\documentclass{ametsoc_draftnolineno}

\journal{jas}

\nolinenumbers

\newcommand{\pf}[2]{\ensuremath{\frac{\partial{#1}}{\partial{#2}}}}

\newcommand{\f}[2]{\ensuremath{\frac{#1}{#2}}}

 \mathchardef\mhyphen="2D

\definecolor{myred}{RGB}{163, 20, 47}
\definecolor{myblue}{RGB}{0, 115,190}
\definecolor{mypurp}{RGB}{127,47,143}
\definecolor{mygold}{RGB}{239,178,32}
\definecolor{mygreen}{RGB}{0,129,0}
\definecolor{myorang}{RGB}{218,84,25}

\makeatletter
\newif\ifpreprintoption
\@ifclasswith{ametsoc}{twocol}{\preprintoptiontrue}{\preprintoptionfalse}
\makeatother

\bibpunct{(}{)}{;}{a}{}{,}

\title{Scaling Behavior of a Turbulent Kinetic Energy Closure Scheme for the Stably Stratified Atmosphere: A Steady-State Analysis}

    \authors{
Michael MacDonald\correspondingauthor{\nolinenumbers Michael MacDonald,  Department of Mechanical Engineering, The University of Auckland, Private Bag 92019, Auckland Mail Centre, Auckland 1142, New Zealand}\thanks{\nolinenumbers Current affiliation: Department of Mechanical Engineering, The University of Auckland, Auckland, New Zealand}
 and    Jo{\~{a}}o Teixeira
}
      	\affiliation{\nolinenumbers Jet Propulsion Laboratory, California Institute of Technology, Pasadena, California, USA\\
\vspace{0.3cm}
      	\normalfont\textcopyright \hspace{0.02cm} 2020 California Institute of Technology. U.S.~Government sponsorship acknowledged.}
\email{michael.macdonald@auckland.ac.nz}

\abstract{
We present a turbulent kinetic energy (TKE) closure scheme for the stably stratified atmosphere in which the mixing lengths for momentum and heat are not parameterized in the same manner. The key difference is that, while the mixing length for heat
tends towards the stability independent  mixing length for momentum in neutrally stratified conditions, it tends towards one based on the Brunt--V{\"a}is{\"a}l{\"a} time scale and square root of the TKE in the limit of large stability. This enables a unique  steady-state solution for TKE to be obtained, which we demonstrate would otherwise be impossible if the mixing lengths were the same. 
Despite the model's relative simplicity, it is shown to perform reasonably well with observational data from
the 1999 Cooperative Atmosphere-Surface Exchange Study (CASES-99) using commonly employed model constants.
Analyzing the scaling behavior of the non-dimensional velocity and potential temperature gradients, or of the stability (correction) functions, reveals that for large stability the present model scales in the same manner as the first-order operational scheme of Viterbo et al.~(Quart.~J.~Roy.~Meteor.~Soc.~{\bf 125}, 2401--2426, 1999). Alternatively, it appears as a blend of two cases of the TKE closure scheme of Baas et al.~(Bound.-Layer Meteor.~{\bf 127}, 17--36, 2008).
Critically, because a unique steady-state TKE can be obtained, the present model   avoids the non-physical behavior identified in one of the cases of \cite{Baas08}.
} 

\begin{document}

%% Necessary!
\maketitle

%%%%%%%%%%%%%%%%%%%%%%%%%%%%%%%%%%%%%%%%%%%%%%%%%%%%%%%%%%%%%%%%%%%%%
% 
%           INTRODUCTION
%
%%%%%%%%%%%%%%%%%%%%%%%%%%%%%%%%%%%%%%%%%%%%%%%%%%%%%%%%%%%%%%%%%%%%%
\section{Introduction}
\label{sect:intro}
The stably stratified planetary boundary layer (SBL) forms when the planetary surface is cooler than the air above,
frequently occurring at night or in polar regions.
The thermal stratification acts to suppress turbulent motions, due to the effort in drawing up heavier (cooler) fluid from below and pulling down lighter (warmer) fluid from above.
The degree of thermal stratification is traditionally classified as either weakly stable, associated with moderate winds and sustained turbulence,
or very stable, associated with weak winds and weak or intermittent turbulence. The very stable case is also sensitive to other factors such as gravity waves, longwave radiation and local topography \citep{Nieuwstadt84,Mahrt99}.

Despite the prevalence of the SBL and decades of research,  it remains particularly challenging for atmospheric models to realistically represent \citep{Viterbo99,Cuxart06}.
Inaccurate representations of the SBL affect the predicted momentum and heat fluxes at the planetary surface which can go on to have significant large-scale influences in both weather and climate forecasting \citep{Cuxart06}. Models therefore need to be accurate and robust, both in terms of their operational performance but also in their ability to faithfully reproduce fundamental physical processes and states.

In general, atmospheric turbulence parameterizations use the gradient-flux approach, in which vertical turbulent fluxes are related to the vertical gradient through an eddy viscosity or diffusivity. For momentum (subscript $m$) and heat (subscript $h$) these are
\begin{eqnarray}
%%%%-\left(\overline{w'u'}\pf{u}{z}+\overline{w'v'}\pf{v}{z}\right)\hspace{-0.2cm} &=&\hspace{-0.2cm} K_m\left[\left(\pf{u}{z}\right)^2+\left(\pf{v}{z}\right)^2\right]
%%%%\label{eqn:eddyflux_m} \\
%-\overline{w'u'}\hspace{-0.2cm} &=&\hspace{-0.2cm} K_m\pf{u}{z}
%\label{eqn:eddyflux_m} \\
%-\overline{w'v'}\hspace{-0.2cm} &=&\hspace{-0.2cm} K_m\pf{v}{z}
%\label{eqn:eddyflux_mv} \\
-\overline{w'u'} = K_m\pf{u}{z}\hspace{-0.4cm}&,& \hspace{0.1cm} -\overline{w'v'}= K_m\pf{v}{z},
\label{eqn:eddyflux_m} \\
-\overline{w'\theta_v} \hspace{-0.15cm}&=&\hspace{-0.15cm}  K_h \pf{\theta_v}{z},
\label{eqn:eddyflux_h}
\end{eqnarray}
where  $u$ and $v$ are the mean horizontal wind components, $w$ is the vertical wind, $\theta_v$ is the mean virtual potential temperature and primes denote fluctuations. 
%$\overline{w'u'}$ and $\overline{w'\theta_v'}$ are then the vertical turbulent  fluxes of momentum and potential temperature, where
The eddy viscosity, $K_m$, and eddy diffusivity, $K_h$, must then be parameterized.

In order to aid the interpretation of common parameterizations for $K_m$ and $K_h$ discussed below, we briefly introduce the local-scaling theory of \cite{Nieuwstadt84}.
%Before detailing the scheme, we briefly introduce the local-scaling theory of \cite{Nieuwstadt84} which will be used to discuss the schemes scaling behavior. 
This theory states that non-dimensional quantities within the stable boundary layer are only functions of $\zeta\equiv z/\Lambda$, where $\Lambda$ is the \emph{local} 
Obukhov length, defined as
\begin{equation}
\Lambda = -\f{\left(\overline{w'u'}^2+\overline{w'v'}^2\right)^{3/4}}{\kappa ({g}/{\Theta_v})\overline{w'\theta_v'}} = \f{u_*^2}{\kappa ({g}/{\Theta_v})\theta_*}, \label{eqn:Lambda}
\end{equation}
where 
$u_*^2 = (\overline{w'u'}^2+\overline{w'v'}^2)^{1/2}$ is the local friction velocity
and
$\theta_*=-\overline{w'\theta_v'}/u_*$ is the local friction potential temperature, both of which vary with height, and $\kappa=0.4$ is the von K{\'a}rm{\'a}n constant.
This local-scaling theory represents a generalization of Monin--Obukhov similarity theory (MOST), wherein MOST is only valid within the near-ground surface layer and uses the \emph{surface} Obukhov length, $L$.
Essentially, local-scaling theory states that the  non-dimensional  velocity and virtual potential temperature gradients, $\phi_{m,h}$, are solely functions of $\zeta$, where
\begin{eqnarray}
\phi_m(\zeta) &=& \f{\kappa z}{u_*}\left[\left(\f{\partial u}{\partial z}\right)^2+\left(\f{\partial v}{\partial z}\right)^2\right]^{1/2} \label{eqn:phim}
\\
\phi_h(\zeta) &=& \f{\kappa z}{\theta_*}\f{\partial \theta_v}{\partial z}. \label{eqn:phih}
\end{eqnarray}
Defining the shear rate $S^2 = (\partial u/\partial z)^2+(\partial v/\partial z)^2$ and Brunt--V{\"a}is{\"a}l{\"a} (BV) frequency $N^2 = \f{g}{\Theta_v}(\partial \theta_v/\partial z)$
enables these non-dimensional gradient functions to be related to the gradient and flux Richardson numbers as
\begin{eqnarray}
Ri_g \equiv \f{N^2}{S^2} = \f{\zeta\phi_h}{\phi_m^2}  \label{eqn:Rig} \\
Ri_f \equiv \f{({g}/{\Theta_v})\theta_*}{u_*S} = \f{\zeta}{\phi_m} \label{eqn:Rif}.
\end{eqnarray}
The turbulent Prandtl number
 is then given as
\begin{equation}
Pr \equiv \f{u_*/S}{({g}/{\Theta_v})\theta_*/N^2} = \f{Ri_g}{Ri_f}= \f{\phi_h}{\phi_m}.  \label{eqn:Prt}
\end{equation}

First-order closure schemes are commonly used in operational forecast models, wherein $K_{m,h}$ is  parameterized as 
$K_{m,h} = l_{m,h}^2 S \hspace{0.05cm}F_{m,h}(Ri_g)$, where 
$l_{m,h}$ is the mixing length for momentum or heat.
The term
$F_{m,h} = 1/(\phi_m \phi_{m,h})$ is a stability (correction) function
that typically reduces with Richardson number to account for the reduction of turbulent motions with increasing stability.
These first-order schemes therefore only require solving the prognostic equations for the mean variables.
However, experimental data for $\phi_{m,h}$ and therefore $F_{m,h}$ show large scatter for high stabilities \cite[see, for example][]{Hogstrom88,Andreas02,Beare06}, as well as suffering from self-correlation issues \citep{Hicks78,Baas06}.
This leads to a wide range of semi-empirical forms of $F_{m,h}$ suggested in the literature for first-order schemes.  
Some impose a critical Richardson number of approximately 0.2 above which the flow relaminarizes and turbulence ceases entirely \cite[e.g.][]{Businger71,Dyer74}.
While this behavior agrees with MOST in the surface layer, it  is contrary to observational data which shows turbulence persisting for $Ri_g>1$ \citep[][and references therein]{Galperin07,Huang13les,Mahrt14}. 
Moreover, the lack of turbulent mixing can lead to runaway cooling and a decoupling between the near-surface state and that higher up in the SBL \citep{Derbyshire99}.
This motivated 
so-called sharp forms for $F_{m,h}$ which enables turbulence to remain non-negligible for $Ri_g\rightarrow\infty$ \cite[e.g.][]{King01}.
Operational models often require even more mixing than these forms of $F_{m,h}$ provide to avoid the decoupling behavior, leading to much larger $F_{m,h}$ functions that are tuned based on model performance \cite[e.g.][]{Louis82,Beljaars91,Viterbo99}.

So-called 1.5-order closure schemes are more advanced than first-order schemes as they also solve the prognostic turbulent kinetic energy (TKE) equation. This enables $K_{m,h}$ to be parameterized on TKE \cite[e.g.][and others]{Mellor82,Teixeira04} and have been shown to perform well compared to first-order schemes \citep{Cuxart06}.
While popular in research and mesocale models, these 1.5-order schemes are not typically used in operational global atmospheric models \citep{Cuxart06}
and have not received as much attention in terms of their high-level scaling behavior \citep{Baas08}.
The parameterization used is often of the form $K_{m,h} = l_{m,h}\sqrt{e}F_{m,h}$, where $e$ is the TKE and a commonly employed mixing length is
 $l_{m,h} \propto\tau \sqrt{e}$ \citep[][]{Deardorff80,Cuxart00,Cuxart06,Baas08}.
Here, $\tau$ is a timescale typically related to the BV frequency in stably stratified flows \citep{Deardorff80}.

Critically, 
as will be demonstrated later, 
this commonly employed mixing length parameterization does not yield a unique steady-state solution for TKE,
when the TKE equation consisting solely of shear production, buoyant destruction and dissipation is considered.
This form of the stationary TKE equation is a physically realizable  state that has been observed in stably stratified homogeneous sheared turbulence \citep{Gerz89,Holt92}, is assumed in Monin--Obukhov similarity theory and even
explicitly exploited to achieve stationary conditions in numerical simulations \citep{Chung12}.
Therefore, the inability to gracefully handle this fundamental steady-state behavior represents a troublesome deficiency in these 1.5-order schemes.
%The deficiency was observed in \cite{Baas08} when particular model constants did not yield a steady-state solution, resulting in a non-physical increase of TKE each time step.

A similar, yet distinct, steady-state deficiency has been identified by \cite{Baumert00} for the 2.5-order scheme of \cite{Mellor82}. This involves, in part, solving prognostic equations for TKE and additionally for a master mixing length, $l$. \cite{Baumert00} showed that this master mixing length prognostic equation is essentially the same as the TKE dissipation rate equation, therefore referring to the scheme as a $k$--$\varepsilon$ closure scheme following engineering nomenclature. At any rate, the form of these two prognostic equations in \cite{Mellor82} yields no steady state solution for TKE, regardless of mixing length parameterization. \cite{Baumert00} speculated that this issue was a possible reason for ad-hoc limiters being applied to the mixing length in future 2.5-order models \citep[e.g.~in][]{Galperin88}. Alternatively, additional terms may be added to the TKE and dissipation equations to represent the transfer of turbulent energy to internal gravity waves, which appears to avoid the issue  \citep{Baumert04,Zeng20}.
Rather than adding complexity and computational cost by solving an additional highly parameterized prognostic equation for dissipation, we aim to retain the simplicity of the present 1.5-order closure scheme by only solving for TKE.

%We will compare the present  TKE closure scheme with the first-order closure models described above in terms of $\phi_{m,h}$ and $F_{m,h}$, similar to the analysis performed in \cite{Baas08} for their TKE closure scheme. Comparisons are also made to the scheme of \cite{Baas08} which used the mixing length formulation $l_{m,h}\propto\tau\sqrt{e}$ described above. In particular, we focus on their cases C and D. Case C had model constants that were their original operational values and had a critical gradient Richardson number of 1.3, above which turbulence ceased. Case D, meanwhile, had no critical $Ri_g$ and used model constants that, for their model, yielded a non-physical scenario where there was no TKE solution in the limit of large stability. 
%The present closure scheme uses different formulations for the mixing lengths for momentum and heat to avoid this non-physical behavior.

The outline of this paper is a follows. The TKE closure model is developed in section \ref{sect:model}, with the steady-state deficiency demonstrated and then addressed in section \ref{sect:model}\ref{subsect:mixingLength}. A brief validation of the model is provided in section \ref{sect:validation}, although as we are more concerned with the broad properties and steady-state behavior of the model we do not attempt to finely tune the model constants. The scaling  behavior of the model is then analyzed in section \ref{sect:scaling}, similar to the analysis by \cite{Baas08} for their TKE closure scheme, wherein comparisons are made to the first- and 1.5-order closure models in terms of $\phi_{m,h}$ and $F_{m,h}$. Conclusions are then offered in section \ref{sect:conc}.

%%%%%%%%%%%%%%%%%%%%%%%%%%%%%%%%%%%%%%%%%%%%%%%%%%%%%%%%%%%%%%%%%%%%%
% 
%          MODEL
%
%%%%%%%%%%%%%%%%%%%%%%%%%%%%%%%%%%%%%%%%%%%%%%%%%%%%%%%%%%%%%%%%%%%%%
\section{TKE closure model}
\label{sect:model}

\subsection{Model formulation}
\label{subsect:formulation}
The prognostic equation for TKE ($e$) assuming horizontally homogeneous conditions can be written as \cite[][]{Stull88}
\begin{eqnarray}
\hspace{-0.25cm}\pf{ e}{t} \hspace{0.0cm}= \hspace{0.0cm}
-\pf{}{z}\left(\overline{w'e}+\f{\overline{w'p'}}{\rho_0}\right)
+\f{g}{\theta_{v0}}\overline{w'\theta_v'}   
\hspace{0cm}\nonumber
\\
\hspace{0cm}
-\left(\overline{w' u'}\pf{u}{z} + \overline{w'v'}\pf{v}{z}\right)                   
-\varepsilon,                                                                 
\label{eqn:tkebudget}
\end{eqnarray}
where the terms on the right-hand side correspond to 
transportation (due to turbulence and pressure diffusion), 
buoyant production, 
shear (or mechanical) production,
and dissipation, 
respectively.

The transport term is herein neglected, similar to other SBL modeling studies \citep[e.g.][]{Ellison57,Zilitinkevich10,Wilson15}, 
as it is often found to be small in the SBL \citep{Nieuwstadt84} and
 typically has negligible impact in models when included in the full TKE prognostic equation \citep{Baas08}.
However, especially during strong intermittent turbulent events, the transport term can become significant \citep{Cuxart02} indicating this assumption may restrict us to sustained turbulent conditions such as in weakly and moderately stable cases.
From (\ref{eqn:tkebudget}), we note that for steady stably stratified turbulence without the transport term, the buoyant destruction cannot exceed the shear production, so that $Ri_f<1$
and should tend to a constant value for large stability (\citealt{MoninYaglom71}, \S7.3; \citealt{Zilitinkevich10}).

%\cites[\S 7.3][]{Monin71,Zilitinkevich10}.
%Following the eddy-diffusivity approach, we parameterize the turbulent fluxes as
%\begin{eqnarray}
%-\left(\overline{w'e}+\f{\overline{w'p'}}{\rho_0}\right) &=& K_m \pf{e}{z} 
%\label{eqn:eddyflux_e} \\
%-\overline{w'\theta_v'} &=& K_h \pf{\theta_v}{z} 
%\label{eqn:eddyflux_h}\\
%-\left(\overline{w'u'}\pf{u}{z}+\overline{w'v'}\pf{v}{z}\right)\hspace{-0.3cm} &=&\hspace{-0.3cm} K_m\left[\left(\pf{u}{z}\right)^2+\left(\pf{v}{z}\right)^2\right]
%\label{eqn:eddyflux_m}
%\end{eqnarray}
%where $K_m$  and $K_h$ are the momentum eddy viscosity and heat eddy diffusivity, respectively, which are in turn parameterized as

Following other SBL closure schemes \citep[e.g.][]{Deardorff80,Mellor82,Cuxart06,Mauritsen07,Baas08}, the dissipation is parameterized using the Kolmogorov approach, with
% with $K_{m,h} = l_{m,h}^2 S \hspace{0.05cm}f_{m,h}(Ri_g)$, where $l_{m,h}$ is a mixing length, $S^2 = (\partial u/\partial z)^2+(\partial v/\partial z)^2$ the shear rate, and 
% simplest and most commonly used approach in operational schemes
 \begin{equation}
 \varepsilon = C_{\varepsilon}\f{e^{3/2}}{l_\varepsilon},
 \end{equation}
where $l_\varepsilon$ is the dissipation length scale and $C_\varepsilon$ a constant.
Under neutrally stratified, isotropic and homogeneous turbulence, $C_\varepsilon$ is often taken to be 0.7, however these assumptions of isotropy and homogeneity break down with increasing stratification. The value of $C_\varepsilon$ is therefore often reduced as a result \citep{Cuxart06}; here we use $C_\varepsilon=0.16$ from \cite{Teixeira04} which is similar to the stable-limit value of 0.19 in \cite{Deardorff80}. As with neglecting the TKE transport term above, this dissipation parameterization may not be suitable for very stable and intermittent cases when the turbulence may be anisotropic and inhomogeneous.

Following the eddy-diffusivity approach (\ref{eqn:eddyflux_m}--\ref{eqn:eddyflux_h}), we parameterize the momentum eddy viscosity and heat eddy diffusivity on TKE, with
 \begin{eqnarray}
K_m &=& C_m l_m \sqrt{e} 
\label{eqn:Km}
\\
K_h &=& C_h l_h\sqrt{e},
\label{eqn:Kh}
 \end{eqnarray}
thus requiring the mixing lengths for momentum and heat, $l_m$ and $l_h$ to be determined.
% The coefficients $C_m$ and $C_h$ are constants, taken here to be $C_m=0.1$ \cite[e.g.][]{Deardorff80}
%and $C_h=C_m/0.75$.
The coefficients $C_m$ and $C_h$ are taken to be constants, with $C_m=0.1$ \citep[from the large eddy simulation TKE closure scheme of ][]{Deardorff80} and $C_h=C_m/0.75$. This will be shown to yield a turbulent Prandtl number in neutrally stratified conditions of 0.75.
Finally, under steady state conditions and with the assumptions made above, (\ref{eqn:tkebudget}) then becomes
 \begin{equation}
 e = \f{l_\varepsilon}{C_{\varepsilon}} C_m l_m S^2 \left(1-\f{C_h}{C_m}\f{l_h}{l_m}Ri_g\right).
 \label{eqn:tkesteady}
 \end{equation}
 %where $S^2 = (\partial u/\partial z)^2 + (\partial v/\partial z)^2$ is the shear rate and 
% \begin{equation}
% Ri_g = \f{\beta\pf{\theta_v}{z}}{\left(\pf{u}{z}\right)^2+\left(\pf{v}{z}\right)^2}=\f{N^2}{S^2},
% \end{equation}
%is the gradient Richardson number, $N$ being the Brunt--V{\"a}is{\"a}l{\"a} frequency with coefficient of thermal expansion $\beta=g/\theta_{v0}$.

%%%%%%%%%%%%%%%%%%%%%%%%%%%%%%%%%%%%%%%%%%%%%%%%%%%%%%%%%%%%%%%%%%%%%
% 
%          MIXING LENGTHS
%
%%%%%%%%%%%%%%%%%%%%%%%%%%%%%%%%%%%%%%%%%%%%%%%%%%%%%%%%%%%%%%%%%%%%%
\subsection{Mixing length definitions}
\label{subsect:mixingLength}
For the mixing length parameterization, 
a common approach \citep[e.g.][]{Deardorff80,Lenderink04,Cuxart06,Baas08} is to put $l_{m,h,\varepsilon}=\tau\sqrt{e}$, where $\tau$ is some timescale typically related to the inverse of the BV frequency \citep{Deardorff80}.
However, this is problematic as we see that (\ref{eqn:tkesteady})  then results in
\begin{equation}
\label{eqn:notke}
e = \tau^2 e \f{C_m}{C_\varepsilon} S^2\left(1-\f{C_h}{C_m}Ri_g\right),
\end{equation}
which does not yield a unique solution for TKE.
%The assumptions and parameterizations made in section \ref{sect:model}\ref{subsect:formulation} to reach this stage are fairly standard, indicating this flaw under steady-state conditions is widespread to many TKE closure models. 
%As with the similar steady-state deficiency identified by \cite{Baumert00} for the 2.5-order closure scheme in \cite{Mellor82}, we speculate that this deficiency is a possible reason that additional ad-hoc limits on the mixing lengths are imposed in TKE closure models.
%Instead, (\ref{eqn:notke}) is sometimes used
%to  give the critical gradient Richardson number, $Ri_{gc}$, as a function of model constants  \citep[e.g.][]{Cuxart00,Baas08}.
% Here, it would be $Ri_{gc}=C_m/(C_\varepsilon+C_h)$ (if $\tau=1/N$), above which turbulence ceases.
This could readily result in atmospheric models predicting the unbounded increase of TKE with time, as was observed in \cite{Baas08} for particular model constants.
We therefore speculate that this issue perhaps contributes to the numerical instabilities often associated with 1.5-order closure schemes \citep{Lenderink04comp}.
Furthermore, like the similar steady-state issue identified with 2.5-order models in \cite{Baumert04}, 
the non-uniqueness may be the reason that additional ad hoc limiters on mixing lengths, or even the TKE, are used to artificially constrain 1.5-order models.

%\textcolor{red}{Comment on implication (see red text above in introduction). Does this cause/appear as a numerical stability issue?}
%Cuxart00 - equation 26
%Baas08 - equation 20
%\textcolor{change}{
%Additional ad hoc limiters on mixing lengths and model coefficients may therefore be needed \textcolor{red}{[can we speculate they are used?]} to artificially constrain the model due to the %steady-state deficiency. Here, we aim to resolve this issue without resorting to these limiters and constraints.
%}
%\textcolor{red}{
%\bf{Limiters used in \cite{Lenderink04} for mixing length ($l_{m,h} = (l_{min}{^-1}+l_s^{-1})^{-1}$ where $l_s = \tau \sqrt{e}$ and $l_{min}$ = Blackader formulation, so for very stable %$l_{m,h}\rightarrow l_{min}$ as $l_s\rightarrow0$.}
%\bf{Numerical stability issues highlighted in \cite{Lenderink04comp}}
%}
%\textcolor{red}{Joao - is this issue a possible reason why TKE schemes aren't commonly used (ignoring the additional computation required)? Is there any talk in the literature about it? Could it %have been (inadvertently) attributed to a numerical stability issue (as any tendency toward steady-state could result in the solution blowing up - see \S6.2 of \cite{Baas08})?
%Are limiters/artificial constraints common in TKE schemes, perhaps due to this issue?
%Presumably any tendency towards steady state in the model could result in the solution blowing up without some kind of ad-hoc limters (see other red text below).
%}

The non-uniqueness of (\ref{eqn:notke}) emerges due to the identical mixing length parameterization employed. To resolve this, we instead use different mixing length formulations for momentum and heat, a similar approach as done in \cite{Teixeira04imp} for unstably stratified flows. 
Given that the mixing length is a highly conceptual parameter representing the length scale of turbulent mixing, then there is no physical reason that the two must be parameterized in the same manner.
We also aim to avoid the use of any ad hoc limiters or stability correction functions on the mixing length or on TKE.

For momentum we simply use the formulation of \cite{Blackadar62},
\begin{equation}
\f{1}{l_m}=\f{1}{\kappa z}+\f{1}{l_\infty},
\label{eqn:lm}
\end{equation}
which scales as $\kappa z$ close to the surface and
where $l_\infty$ is the asymptotic turbulent mixing length far from the surface. 
Typically $l_\infty$ is approximately 40 to 200 m \citep{Cuxart06}, however
 studies using LES \citep{Huang13model} and field observations \citep{Kim92,Sun11}
have suggested smaller values of approximately 5 to 15 m for stably stratified flows.
Here we use $l_\infty=7$ m following \cite{Huang13model}.

Note that while there is no dependence on stability, unlike some other momentum mixing lengths formulations for first-order schemes \citep{Delage74,Huang13model}, in 1.5-order closure schemes the  eddy viscosity will be indirectly affected by stability due to its dependence on $e$ through (\ref{eqn:Km}).

To distinguish the parameterization for the mixing length for heat from that of momentum, we can take advantage of the observed increase in the turbulent Prandtl number, $K_m/K_h$, with stability
\citep{Ellison57,MoninYaglom71,Kim92,Sukoriansky06,VS10,Huang13les,Li19}.
This increase is often attributed to momentum being mixed more efficiently than heat due to gravity waves  \citep{Lenderink04,Anderson09}.
Ideally the gravity waves and turbulence would be parameterized separately, where attempts have been made to do so with higher-order closure schemes \citep{Zilitinkevich02} or by adding additional source or sink terms to the TKE  and dissipation equations in $k$--$\varepsilon$ schemes \citep{Baumert04,Zeng20}. However this is challenging due to the difficulty in even distinguishing the two motions of turbulence and gravity waves apart from measurement data \citep{Stewart69,Jacobitz05}. Therefore, in the present 1.5-order closure scheme we will simply require the turbulent Prandtl number to become very large for extremely stable situations ($Ri_g\rightarrow\infty$), to approximate this influence of gravity waves \citep{Lenderink04}.
Note that, by requiring an unbounded increase in $Pr$ with no critical $Ri_g$, this scheme will not obey the stable-limit local-scaling theory of \cite{Nieuwstadt84} which is based on the $z$-less scaling arguments of \citep{Wyngaard72}.
%these theories assume all dimensionless parameters must tend towards finite values in the stable limit.
Meanwhile, the mixing length for heat under neutrally stratified conditions ($Ri_g\rightarrow0$) should converge toward that of momentum.

The two limiting behaviors above can be achieved with
\begin{equation}
\f{1}{l_h}=\f{1}{\tau \sqrt{e}}+\f{1}{l_m},
\end{equation}
or equivalently
\begin{equation}
l_h = \f{l_m \tau \sqrt{e}}{\l_m+\tau \sqrt{e}},
\label{eqn:lh}
\end{equation}
where $\tau=\alpha/N$ is a time scale based on the BV frequency and constant $\alpha=0.76$ \citep{Deardorff80,Moeng84}.
We see that,
for extremely stable situations as $\tau\sqrt{e}/l_m\rightarrow0$, the mixing length for heat $l_h\rightarrow \tau\sqrt{e}$ and the turbulent Prandtl number $K_m/K_h\rightarrow\infty$.
Meanwhile,
under neutrally stratified conditions as $\tau\sqrt{e}/l_m\rightarrow\infty$, we have $l_h\rightarrow l_m$ and $K_m/K_h\rightarrow C_m/C_h=0.75$,
as required.

Finally, we set the dissipation mixing length $l_{\varepsilon}=\mu l_h$ with
$C_\varepsilon/\mu=0.08$, a value similar to that used in other studies of the SBL \citep[e.g.][]{Cuxart00,Lenderink04,Baas08}.
Conventionally the dissipation mixing length is based on that of momentum, albeit in schemes where there is no distinction between $l_m$ and $l_h$ such that $l_\varepsilon \propto \tau \sqrt{e}$.
Here, we use the mixing length for heat as it retains this dependency on TKE in the stable limit and is therefore similar to these previous schemes.
Using a blend of $l_m$ and $l_h$ to define the dissipation mixing length \citep[e.g.~equation (8) of][]{Teixeira04imp} does not significantly change the results or conclusions of the present study, presumably because $l_h$ (and hence $l_\varepsilon$) already has some dependency on $l_m$.

Under the above mixing length formulation, (\ref{eqn:tkesteady})  then becomes
 \begin{equation}
 e = \f{\mu}{C_{\varepsilon}} C_m l_m S^2 \f{l_m \tau \sqrt{e}}{\l_m + \tau\sqrt{e}}\left(1-\f{C_h}{C_m}\f{\tau\sqrt{e}}{l_m+\tau\sqrt{e}}Ri_g\right),
 \label{eqn:tkesteadysub}
 \end{equation}
and can now be numerically solved for the steady-state turbulent kinetic energy.

% {\bf NOW WHAT}
% and is thus solvable for $e$. 
%We set $C_m=0.1$, $C_h=C_m/0.85$ so that at neutral conditions the turbulent Prandtl number is 0.85 and $C_\varepsilon/\mu=0.08$.

%%%%%%%%%%%%%%%%%%%%%%%%%%%%%%%%%%%%%%%%%%%%%%%%%%%%%%%%%%%%%%%%%%%%%
% 
%          VALIDATION
%
%%%%%%%%%%%%%%%%%%%%%%%%%%%%%%%%%%%%%%%%%%%%%%%%%%%%%%%%%%%%%%%%%%%%%
\section{Model validation with CASES-99}
\label{sect:validation}
Before analyzing the scaling properties of the model, we provide a short validation that the model agrees reasonably well with field experiments. 
For this purpose, we use the 1999 Cooperative Atmosphere-Surface Exchange Study \cite[CASES-99;][]{Poulos02}, a field campaign conducted in
southeastern Kansas (37.65$^\circ$N, 96.74$^\circ$W; 440 m a.s.l.) in October 1999.
Both weakly and very stable conditions were observed, along with other SBL events such as intenal gravity waves.
A sixth-order polynomial fit to the 60-m main tower wind speed and temperature data is used to determine velocity and potential temperature gradients \citep{Sorbjan10}.
The 5-minute averaged measurements are transformed to 1-hour averages and bin-averaged for $Ri_g$, as in \cite{Wilson15}.

%Additionally, we ignore non-steady cases, when the 1-hour averaged $\partial e/\partial t\ge 2.8\times 10^{-5}$ m$^2$ s$^{-3}$;
% this filtering removes approximately 20--30\% of the stably stratified data.
%For reference, the tendency term is approximately $5\times 10^{-5}$ m$^2$ s$^{-3}$ for surface-layer air
%experiencing a strong diurnal cycle, averaged over 6 hours \citep{Stull88}. However, the 1-hour averaging already filters much
%of the non-stationary cases and including or excluding these data do not significantly alter the $Ri_g$-bin averaged data presented.

\setlength{\unitlength}{1cm}
\begin{figure}
\centering
\nolinenumbers
%  \includegraphics[width=19pc,trim = 0 15 0 0,clip=true]{./Figures/CASES99_lm_z50_ST0p100fillstd.eps}
%  \put(-7.9,5.05){(a)}
%  \\
%  \vspace{-0.25cm}
%  \includegraphics[width=19pc]{./Figures/CASES99_lh_z50_ST0p100fillstd.eps}
% \put(-7.9,5.5){(b)}
%  \vspace{-0.25cm}
\includegraphics[width=19pc]{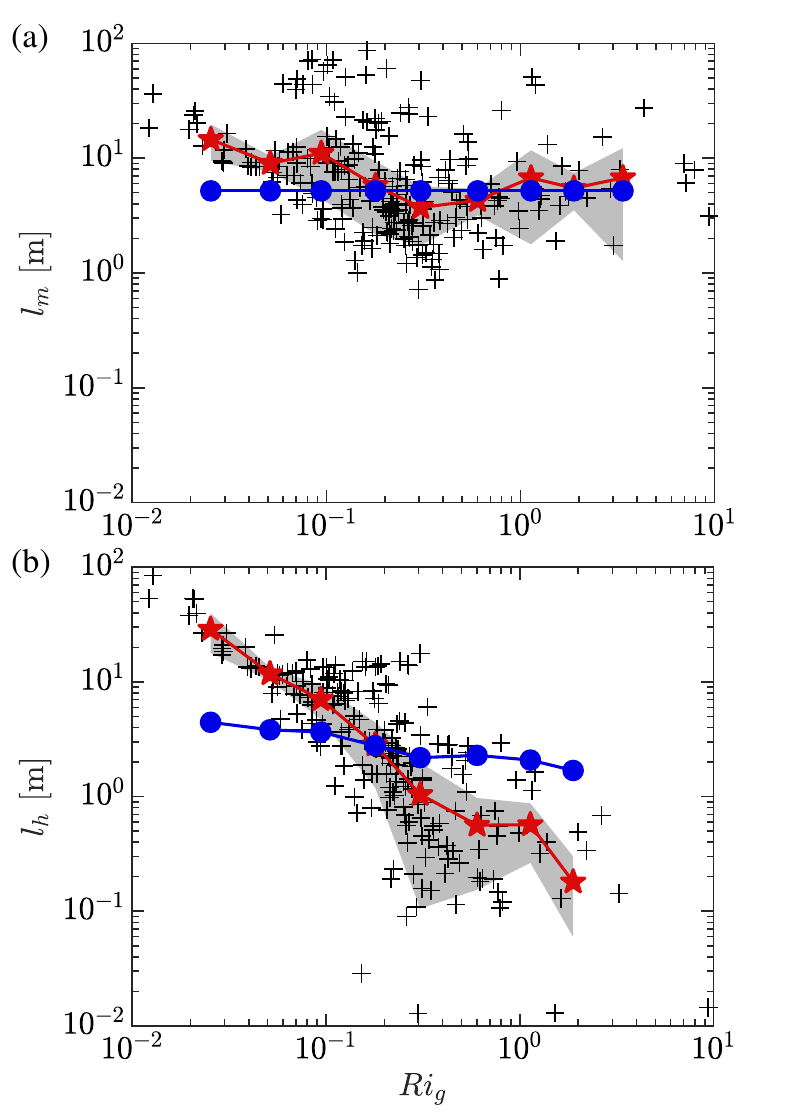}
  \vspace{-0.65cm}
\nolinenumbers
\caption{
(a) Momentum 
and 
(b) heat 
mixing lengths 
at 
 $z=50$ m,
as a function of gradient Richardson number, $Ri_g$.
Black symbols are 1-hour averaged CASES-99 data; 
red stars are the corresponding $Ri_g$-bin averaged data,
where we have assumed (\ref{eqn:eddyflux_m}--\ref{eqn:eddyflux_h}) and (\ref{eqn:Km}--\ref{eqn:Kh}) to determine $l_{m,h}$.
Gray shading indicates $\pm 1$ standard deviation of the $Ri_g$-bin averaged data.
Blue circles are the mixing lengths of (\ref{eqn:lm}) and (\ref{eqn:lh}), using the TKE from CASES-99.
}
\label{fig:mixingLengths}
\end{figure}

Figure \ref{fig:mixingLengths} shows the mixing lengths for momentum and heat for the CASES-99 data at $z=50$ m.
Here, we assume that the shear production of TKE is equal to $K_m S^2$ and use  (\ref{eqn:eddyflux_h}) to compute $K_m$ and $K_h$, respectively.
The mixing lengths, $l_{m,h}$, are then obtained assuming  the TKE eddy diffusivity parameterization given by (\ref{eqn:Km}--\ref{eqn:Kh}).
The $Ri_g$-bin averaged mixing lengths from CASES-99 are shown in red, while the blue symbols show the mixing lengths we would obtain using the present formulation given by (\ref{eqn:lm}) and (\ref{eqn:lh}), where in (\ref{eqn:lh}) we use the TKE from CASES-99.
We see that $l_m$ (Fig.~\ref{fig:mixingLengths}a) is not particularly sensitive to $Ri_g$, thus justifying the use of a stability independent mixing length for momentum (\ref{eqn:lm}).
As mentioned in section \ref{sect:model}\ref{subsect:mixingLength}, this is because within 1.5-order closure schemes the eddy viscosity remains dependent on stability through $e$ in (\ref{eqn:Km}), even if $l_m$ is stability independent. This would not be the case for first-order models which therefore often incorporate the stability correction function $F_m$, or mixing lengths defined to be functions of $Ri_g$ \citep[e.g.][]{Huang13model}.

The mixing length for heat (Fig.~\ref{fig:mixingLengths}b), meanwhile, shows a much stronger dependence on $Ri_g$. This is
somewhat captured by the present $l_h$ formulation ((\ref{eqn:lh}), blue line) and could be improved by reducing $\alpha$ in the time scale $\tau=\alpha/N$ and adjusting other constants accordingly.
As we are more interested in the broad properties of the present parameterization, we will continue to use the present model constants that are commonly used in the literature and
not attempt to finely tune the constants.

\setlength{\unitlength}{1cm}
\begin{figure}
\centering
%  \includegraphics[width=19pc,trim = 0 15 0 0,clip=true]{./Figures/CASES99_Prt_z50_ST0p100fillstd.eps}
%  \put(-7.9,5.05){(a)}
%  \\
%  \vspace{-0.25cm}
%  \includegraphics[width=19pc]{./Figures/CASES99_e_z50_ST0p100fillstd.eps}
%  \put(-7.9,5.5){(b)}
%  \vspace{-0.25cm}
\includegraphics[width=19pc]{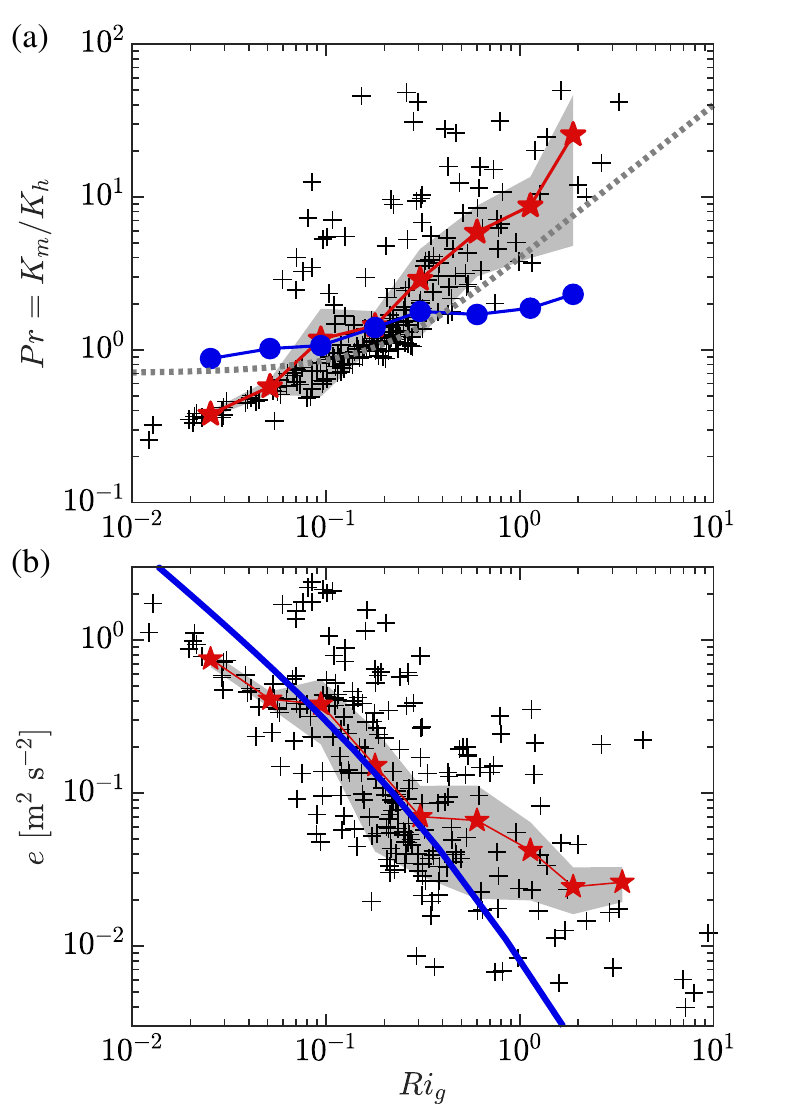}
  \vspace{-0.65cm}
\caption{
(a) Turbulent Prandtl number, $K_m/K_h$, 
and
(b) turbulent kinetic energy, $e$,
at 
$z=50$ m,
as a function of gradient Richardson number, $Ri_g$.
Symbols same as Fig.~\ref{fig:mixingLengths}.
Gray dotted line in (a) is the prediction of \cite{VS10};
blue solid line in (b) comes from solving (\ref{eqn:tkesteadysub}).
}
\label{fig:turbPrKE}
\end{figure}

Figure \ref{fig:turbPrKE}(a) shows the turbulent Prandtl number.
The CASES-99 data show a clear increase with $Ri_g$,
which is again reasonably captured by the present formulation. These are
in agreement with the $Pr$ formulation proposed by \citeauthor{VS10} (\citeyear{VS10}, gray dashed line).
Finally, Fig.~\ref{fig:turbPrKE}(b) shows the TKE from the CASES-99 data along with that obtained from solving the present parameterized steady-state TKE equation of (\ref{eqn:tkesteadysub}).
The lapse rate in the model is fixed to the CASES-99 average at $z=50$ m of $d\theta/dz=40$ K/km,
such that the time scale $\tau\approx 20.1$ s,
and the shear rate (and thus $Ri_g$) is varied.
The resulting TKE predicted by the model shows good agreement with the CASES-99 data and correctly reduces with $Ri_g$.

 This analysis can be repeated at different heights where data are available from the CASES-99 main tower and leads to similar results (not shown).
Ultimately, the above validation demonstrates that the present TKE model with separately parameterized mixing lengths for momentum and heat
captures the essential behavior of the CASES-99 field data.
This is despite the model being relatively simplistic, in which only shear production, buoyant destruction and dissipation of the TKE assuming steady state conditions are considered and standard model constants are employed.
% The mixing length parameterization 

%%%%%%%%%%%%%%%%%%%%%%%%%%%%%%%%%%%%%%%%%%%%%%%%%%%%%%%%%%%%%%%%%%%%%
% 
%          SCALING
%
%%%%%%%%%%%%%%%%%%%%%%%%%%%%%%%%%%%%%%%%%%%%%%%%%%%%%%%%%%%%%%%%%%%%%

\section{Scaling Behavior}
\label{sect:scaling}

We now look at the scaling behavior of the model in terms of the non-dimensional gradients, $\phi_{m,h}$ (\ref{eqn:phim}--\ref{eqn:phih}), and stability functions, $F_{m,h}=1/(\phi_m \phi_{m,h})$,  that are often employed in first-order closure schemes.
As the parameterized steady state TKE equation without transport  term (\ref{eqn:tkesteadysub}) is independent of $z$ 
%(apart from in $l_m$, where we use $z=50$ m following the CASES-99 data validation above) 
we must determine
an appropriate vertical length scale. This is achieved by noting that for neutrally stratified surface layer flows $\phi_m(\zeta\rightarrow0)=1$, which therefore prescribes $z$ when $\kappa=0.4$ is already specified \citep{Chung12}. This choice of $z$ simply guarantees $\phi_m(\zeta=0)=F_m(Ri_g=0)=1$.

\setlength{\unitlength}{1cm}
\begin{figure}
\centering
%  \includegraphics[width=19pc,trim = 0 17 0 0,clip=true]{./Figures/Model_phi_m2014.eps}
%  \put(-7.9,5.05){(a)}
%  \\
%  \vspace{-0.25cm}
%  \includegraphics[width=19pc]{./Figures/Model_phi_h2014.eps}
%  \put(-7.9,5.5){(b)}
%  \vspace{-0.25cm}
\includegraphics[width=19pc]{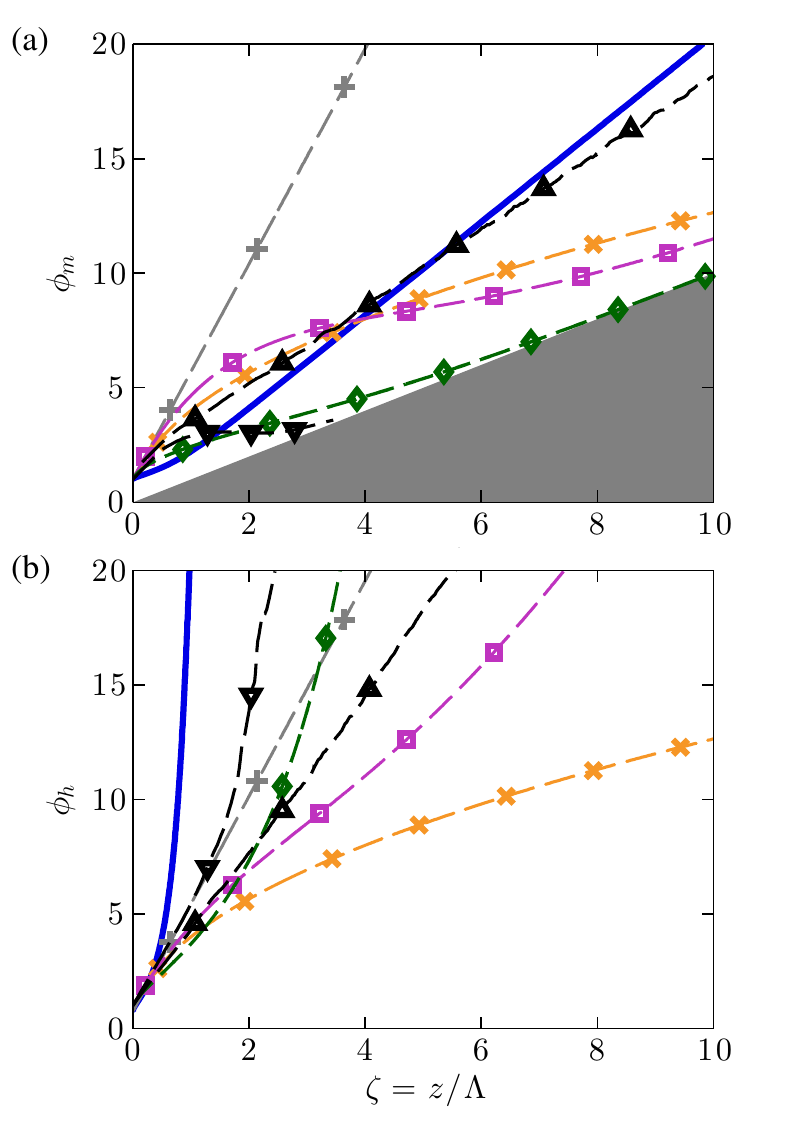}
  \vspace{-0.65cm}
\caption{
Non-dimensional  (a) velocity gradient, $\phi_m$ (\ref{eqn:phim})
and 
(b)
potential temperature gradient, $\phi_h$ (\ref{eqn:phih})
 against non-dimensional vertical position,
$z/\Lambda$.
Line styles:
blue, present model;
gray plus, \cite{Businger71} and \cite{Dyer74};
orange cross, \cite{King01};
purple squares, \cite{Beljaars91};
green diamonds, \cite{Viterbo99};
%black left, up and down triangles, cases A, C and D of \cite{Baas08}, respectively.
black up and down triangles, cases C and D of \cite{Baas08}, respectively.
Gray triangular area in (a) shows the non-physical $Ri_f>1$ regime.
}
\label{fig:phimh}
\end{figure}

Figure \ref{fig:phimh} shows the non-dimensional velocity gradient, $\phi_m$, and potential temperature gradient, $\phi_h$, for the present model. This is accompanied by a selection of first-order closure models often encountered in the literature, introduced in \S\ref{sect:intro},  as well as cases  C and D of the 1.5-order TKE closure model of \cite{Baas08}.
In \cite{Baas08}, the problematic $l_{m,h}\propto\tau\sqrt{e}$ mixing length formulation  was used where 
case C used original operational model constants and had a critical $Ri_g$ of 1.3, above which turbulence ceased.
Case D had no critical $Ri_g$ and used model constants that, for their scheme, 
yielded no solution for TKE in the stable limit.

%This was, in part, due to the aforementioned issue with setting $l_{m,h,\varepsilon}=\tau \sqrt{e}$ for (\ref{eqn:tkesteady}) which yields no unique TKE, although their Case C 

 For momentum (Fig.~\ref{fig:phimh}a), the present model appears similar to case C of the  TKE closure scheme of \cite{Baas08}.
In the limit of large stability, when $\zeta\rightarrow\infty$ and $l_h\rightarrow\tau\sqrt{e}$, it can be shown that the present model tends towards $\phi_m/\zeta= B_m$ where $B_m=1+(C_\varepsilon/\mu)/(C_h\alpha^2)\approx2.04$ is just a function of model constants.
From (\ref{eqn:Rif}), this therefore corresponds to a critical flux Richardson number $Ri_{f,crit}=1/B_m\approx0.49$, similar to the critical value of 0.55 for Case C of \cite{Baas08}. Note that, for positive model constants, the present model guarantees that $Ri_{f,crit}<1$, as required.
Moreover, since the model has a critical $Ri_{f,crit}$ but also  $l_h$ defined such that the turbulent Prandtl number $K_m/K_h\rightarrow\infty$ in the stable limit,
then from (\ref{eqn:Prt}) there can be no critical gradient Richardson number.
The present model limiting behavior for $\phi_m$ is less than
the model of \cite{Businger71} and \cite{Dyer74}, which obeys MOST and where in the stable limit $\phi_m/\zeta=4.7$.
Moreover, it is greater than
  the high-mixing operational schemes of \cite{Beljaars91} and \cite{Viterbo99}, where in the stable limit $\phi_m/\zeta =1$ and approaches the $Ri_f>1$ limiting regime shown by the gray triangular area.

The non-dimensional temperature gradient, $\phi_h$ (Fig.~\ref{fig:phimh}b), meanwhile, shows a very sharp increase with stability for the present model. This is even more rapid than the non-physical Case D of \cite{Baas08}, however the present scheme with separate momentum and heat mixing length formulations avoids the non-physical behavior of that particular case. 
Due to the present model having a critical flux Richardson number but no critical gradient Richardson number, then we see from (\ref{eqn:Rig}) that in the stable limit as $\zeta\rightarrow\infty$, $\phi_h/\zeta=(\phi_m/\zeta)^2Ri_g=(1/Ri_{f,crit}^2)Ri_g$ and thus increases with $Ri_g$. 
From the parameterized TKE equation (\ref{eqn:tkesteadysub}), the TKE in the stable limit (when $l_h\rightarrow\tau\sqrt{e}$) is $e_{stable}= B_e^2 \cdot l_m^2 S^2/Ri_g$ with $B_e= C_m/(C_h B_m\alpha)\approx0.48$. 
Using (\ref{eqn:Lambda}) and (\ref{eqn:Km}--\ref{eqn:Kh}) we can then obtain
%$\Lambda_{stable}=B_\Lambda Ri_g^{-1/4}$ from Eqs.~(\ref{eqn:Lambda}) and (\ref{eqn:Km}--\ref{eqn:Kh}), where $B_\Lambda = (B_m/(C_h \alpha))^{1/2} C_m l_m/\kappa\approx5.81$. 
%$\zeta_{stable}=B_\zeta Ri_g^{1/4}$, where $B_\zeta=\kappa z/(C_m l_m)\cdot(C_h\alpha/B_m)^{1/2}$.  %\approx0.75
$Ri_{g}= (\zeta/B_\zeta)^4$, where $B_\zeta=(C_h\alpha/B_m)^{1/2}\kappa z/(C_m l_m)$.
Ultimately this shows that $\phi_h/\zeta=\zeta^4/(Ri_{f,crit}^2B_\zeta^4)$ in the stable limit, wherein the large exponent of 4 explains the rapid increase of $\phi_h$ with $\zeta$.
%The lack of a finite $\phi_h/\zeta$ in the stable limit (or equivalently no critical $Ri_g$) implies the present scheme does not satisfy the assumptions of Nieuwstadt's (\citeyear{Nieuwstadt84}) local-scaling theory. 

%{\bf Disparity in $\phi_m$ and $\phi_h$: As $Pr$ increases with $Ri_g$ (by design of $l_h$) and $\phi_m/\zeta\rightarrow const$, $\phi_h/\zeta$ must also increase with $Ri_g$. I think \cite{Zilitinkevich10} makes mention of this.}

\setlength{\unitlength}{1cm}
\begin{figure}
\centering
%  \includegraphics[width=19pc,trim = 0 17 0 0,clip=true]{./Figures/Model_F_m2014.eps}
%  \put(-7.85,5.05){(a)}
%  \\
%  \vspace{-0.25cm}
%  \includegraphics[width=19pc,trim = 0 0 0 0,clip=true]{./Figures/Model_F_h2014.eps}
%  \put(-7.85,5.5){(b)}
%  \vspace{-0.25cm}
\includegraphics[width=19pc]{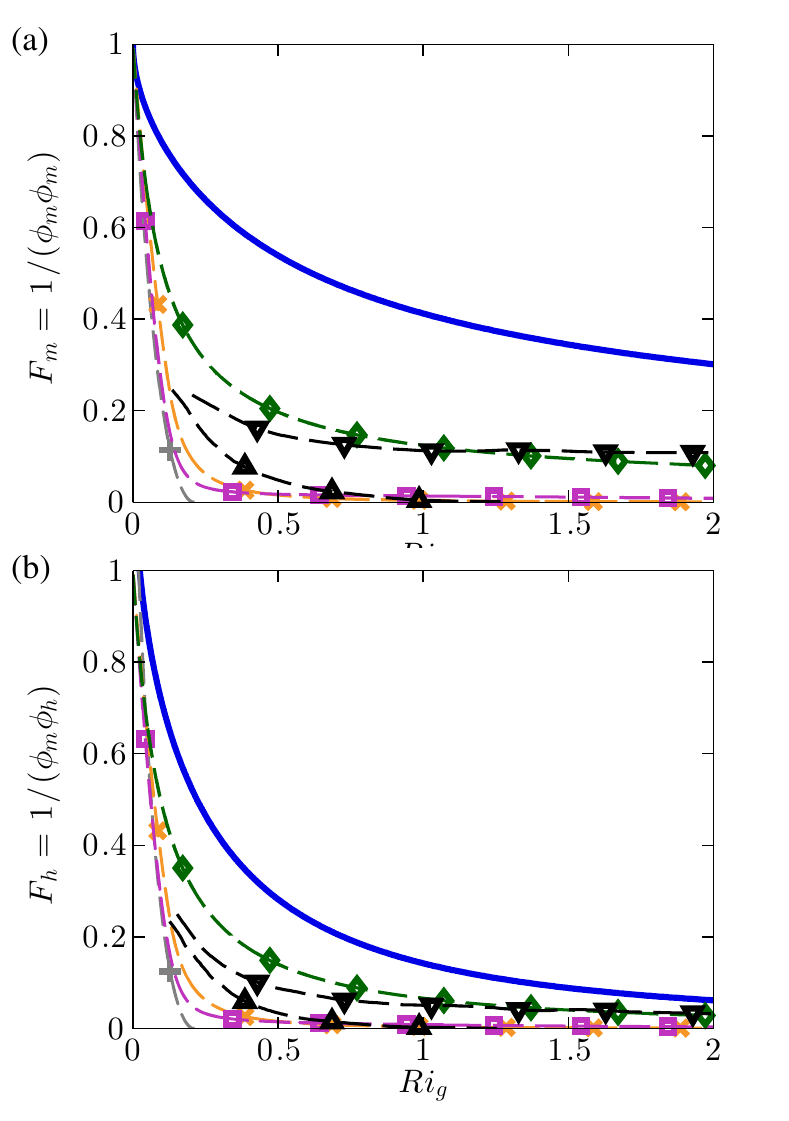}
  \vspace{-0.65cm}
\caption{
(a) Momentum
and
(b) heat
stability functions against gradient Richardson number.
%(c) Turbulent Prandtl number, or ratio of $F_m$ and $F_h$.
Line styles same as Fig.~\ref{fig:phimh}.
%Grey dashed line in (c) is the prediction of \cite{VS10}.
}
\label{fig:Fmh}
\end{figure}

The stability functions for momentum and heat are shown in Fig.~\ref{fig:Fmh}. Both of these functions for the present model are much larger than those of the other schemes, which suggests significantly enhanced mixing. However, this interpretation of enhanced mixing comes from the fact that, in first-order models, $F_{m,h}$ directly modifies $K_{m,h}$ as stability correction functions. For the present closure scheme, $F_{m,h}$ is determined diagnostically from (\ref{eqn:phim}--\ref{eqn:phih}) using the TKE derived from solving (\ref{eqn:tkesteadysub}), which only considers shear production, buoyant destruction and dissipation of TKE.
No additional limiters or correction functions are required in specifying the mixing lengths, eddy viscosity or eddy diffusivity.
 These differences may explain why the steady-state TKE determined by the present model compare reasonably well with observations (Fig.~\ref{fig:turbPrKE}b), despite $F_{m}$ suggesting substantial mixing. 
%{\bf Not sure about this. Does this invalidate my comparisons between first-order and 1.5-order $F_m$?}.
%At high stability, the present model scales the same as the unphysical case D of \cite{Baas08}, as well as high mixing operational scheme of \cite{Viterbo99}, which is based on the Louis-Tiedke-Geleyn (LTG) scheme \citep{Louis82}. These have no critical gradient Richardson number, and 

At large stability, the present model stability functions become  $F_m= (Ri_{f,crit}/B_\zeta )^2Ri_g^{-1/2}\approx 0.4331 Ri_g^{-1/2}$ and $F_h=(Ri_{f,crit}^3/B_\zeta^2 )Ri_g^{-3/2}\approx0.21Ri_g^{-3/2}$. This is the same scaling, although with different model constants, as the large stability limit of \cite{Viterbo99}, where $F_m = 0.1 Ri_g^{-1/2}$ and $F_h= 0.0667Ri_g^{-3/2}$.
These functions are based on the Louis-Tiedke-Geleyn (LTG) scheme 
%\cite[][in particular their equation 10]{Louis82}
\citep{Louis82}
 which are ultimately derived from the work of \cite{Ellison57}.
As in the current work, \cite{Ellison57} assumed that the turbulent Prandtl number became unbounded in the stable limit and derived an expression for $Pr$ from the steady-state TKE equation without the transport term
\cite[see also][\S7.4]{MoninYaglom71}.
This suggests that the high-mixing LTG functions are not entirely non-physical as suggested by some authors \cite[e.g.][]{Baas08}, but are simply a consequence of assuming that the eddy diffusivity  becomes negligible relative to the eddy viscosity in the stable limit.

\section{Conclusions}
\label{sect:conc}
TKE closure schemes are a promising avenue for modeling the stably stratified planetary boundary layer, however they have not been as closely analyzed as the more common first-order approaches. For TKE closure schemes, a commonly employed mixing length parameterization involves using a time scale and the square root of the TKE, $l_{m,h,\varepsilon}=\tau \sqrt{e}$ \citep{Deardorff80}. We show that this common formulation yields no unique solution for TKE when considering the steady-state TKE equation consisting of shear production, buoyant destruction and dissipation. This deficiency may result in the TKE becoming numerically unstable during steady-state conditions, and may explain the use of artificial, ad-hoc limiters to constrain the model.

To avoid this steady-state deficiency, the present TKE closure scheme uses separate mixing lengths for momentum and heat, a similar approach as done in \cite{Teixeira04imp} for unstably stratified flows.
We emphasize that the present formulation does not actually introduce any new mixing length parameterizations that have not individually been used in the literature; rather it merges previous parameterizations into a scheme which can faithfully account for steady-state conditions.

We assume that the mixing length for heat tends towards the (stability independent) mixing length for momentum under neutrally stratified conditions, 
and that the turbulent Prandtl number increases with the gradient Richardson number at large stability.
The increase of $Pr$ somewhat approximates the influence of gravity waves, which mix momentum more efficiently than heat \citep{Lenderink04,Anderson09}.
These two conditions are satisfied with (\ref{eqn:lh}) and results in a scheme with no critical $Ri_g$.
%It does however have a critical flux Richardson number of $Ri_{f,crit}=1/B_m\approx0.49$, which is solely dependent on model constants.
The model enables a unique solution for TKE to be obtained from the steady-state TKE prognostic equation without the transport term  (\ref{eqn:tkesteadysub}), as required.
The assumptions, notably that the turbulence is sustained, isotropic and homogeneous and neglecting intermittent effects,
appear limiting at first. However,
comparisons were made with data from the  CASES-99 field campaign, which observed both weakly and very stable conditions as well as SBL phenomena such as internal gravity waves \citep{Poulos02}, and reasonable agreement was found with the model (Fig.\ref{fig:mixingLengths} and \ref{fig:turbPrKE}).

The present model exhibits some similarities to the TKE closure scheme of \cite{Baas08}, which used the deficient $l_{m,h}=\tau\sqrt{e}$ formulation. Figure \ref{fig:phimh} shows that the present $\phi_m$ is similar to their Case C, which was based on model constants using operational values, while the non-dimensional potential temperature gradient, $\phi_h$, increases very rapidly, similar to their Case D. While Case D in \cite{Baas08} was non-physical in that there was no unique TKE solution in the stable limit, the present formulation with separately parameterized mixing lengths effectively blends their two cases together and avoids this non-physical behavior.

Finally, the stability functions, $F_{m,h}$, of the present scheme are shown to scale in the same manner as the first-order closure Louis-Tiedke-Geleyn (LTG) functions of \cite{Louis82} (or the revised form in \cite{Viterbo99}) in the stable limit.
This is due to the present scheme and the LTG functions, based on the work of \cite{Ellison57}, both of which assume an unbounded turbulent Prandtl number in the limit of large stability with no critical $Ri_g$. As noted with regards to Case D in \cite{Baas08}, this assumption therefore does not obey the stable limit behavior of the local-scaling theory \citep{Nieuwstadt84}, which assumes a critical $Ri_g$ ($z$-less scaling).
However, this suggests that the view that the LTG functions artificially enhance mixing and are nonphysical is instead a consequence of the relaxation of the assumption from local-scaling theory that  $Ri_g$ must remain finite.

The steady-state results discussed in this paper suggest that a similar mixing-length approach would also produce more realistic results for the stable boundary layer in the context of utilizing a fully prognostic TKE equation to determine the eddy diffusivity and eddy viscosity coefficients.

%{\bf Physical insights? See Rev 2 Point 7}

%%%%%%%%%%%%%%%%%%%%%%%%%%%%%%%%%%%%%%%%%%%%%%%%%%%%%%%%%%%%%%%%%%%%%
% ACKNOWLEDGMENTS
%%%%%%%%%%%%%%%%%%%%%%%%%%%%%%%%%%%%%%%%%%%%%%%%%%%%%%%%%%%%%%%%%%%%%
\acknowledgments
This research was carried out at the Jet Propulsion Laboratory, California
Institute of Technology, under a contract with the National Aeronautics and Space
Administration.
Parts of this research were supported by the U.S. Department of Energy, Office of Biological and Environmental Research, Earth System Modeling; the NASA MAP Program; the Office of Naval Research, Marine Meteorology Program and the NOAA/CPO MAPP Program.

%%%%%%%%%%%%%%%%%%%%%%%%%%%%%%%%%%%%%%%%%%%%%%%%%%%%%%%%%%%%%%%%%%%%%
% APPENDIXES
%%%%%%%%%%%%%%%%%%%%%%%%%%%%%%%%%%%%%%%%%%%%%%%%%%%%%%%%%%%%%%%%%%%%%

%% If only one appendix, use
%\appendix
%\appendixtitle{Appendix}
%Text

%% If more than one appendix, use \appendix[<letter>], e.g.,
 %\appendix[A] 
%
%\appendixtitle{Title of Appendix}
%
%
%\subsection{Appendix section}
%
%Appendix

%%%%%%%%%%%%%%%%%%%%%%%%%%%%%%%%%%%%%%%%%%%%%%%%%%%%%%%%%%%%%%%%%%%%%
% REFERENCES
%%%%%%%%%%%%%%%%%%%%%%%%%%%%%%%%%%%%%%%%%%%%%%%%%%%%%%%%%%%%%%%%%%%%%

 \bibliographystyle{ametsoc2014}
 \bibliography{bibliography}

%%%%%%%%%%%%%%%%%%%%%%%%%%%%%%%%%%%%%%%%%%%%%%%%%%%%%%%%%%%%%%%%%%%%%
% TABLES
%%%%%%%%%%%%%%%%%%%%%%%%%%%%%%%%%%%%%%%%%%%%%%%%%%%%%%%%%%%%%%%%%%%%%

%%%%%%%%%%%%%%%%%%%%%%%%%%%%%%%%%%%%%%%%%%%%%%%%%%%%%%%%%%%%%%%%%%%%%
% FIGURES
%%%%%%%%%%%%%%%%%%%%%%%%%%%%%%%%%%%%%%%%%%%%%%%%%%%%%%%%%%%%%%%%%%%%%

\end{document}